\documentclass[review]{elsarticle}

\usepackage{mathptmx}       % selects Times Roman as basic font
\usepackage{helvet}         % selects Helvetica as sans-serif font
\usepackage{courier}        % selects Courier as typewriter font
\usepackage{type1cm}        % activate if the above 3 fonts are
\usepackage{makeidx}         % allows index generation
\usepackage{graphicx}        % standard LaTeX graphics tool
\usepackage{multicol}        % used for the two-column index
\usepackage{url}
\usepackage{color}
\usepackage{ulem}
\usepackage{hyperref}
\usepackage[utf8]{inputenc}
\usepackage{amssymb}

\begin{document}

\begin{frontmatter}

\title{Cross-Referencing Method for Scalable Public Blockchain}

%% Group authors per affiliation:
\author{Takaaki Yanagihara and Akihiro Fujihara}
\address{Department of Information and Communication Systems Engineering, \\
	Chiba Institute of Technology, \\
	2-17-1 Tsudanuma, Narashino, Chiba 275-0016, JAPAN}
\ead{s1522313qq@s.chibakoudai.jp, akihiro.fujihara@p.chibakoudai.jp}

\begin{abstract}
 We previously proposed a cross-referencing method for enabling multiple 
peer-to-peer network domains to manage their own public blockchains and 
periodically exchanging the state of the latest fixed block in the 
blockchain with hysteresis signatures among all the domains via an upper 
network layer. 
 In this study, we evaluated the effectiveness of our method from three 
theoretical viewpoints: decentralization, scalability, and tamper resistance. 
 We show that the performance of the entire system can be improved because 
transactions and blocks are distributed only inside the domain. 
 We argue that the transaction processing capacity will increase to 56,000 
transactions per second, which is as much as that of a VISA credit card system. 
 The capacity is also evaluated by multiplying the number of domains by 
the average reduction in transaction-processing time due to the increase 
in block size and reduction in the block-generation-time interval by domain 
partition. For tamper resistance, each domain has evidence of the hysteresis 
signatures of the other domains in the blockchain. We introduce two types of 
tamper-resistance-improvement ratios as evaluation measures of tamper resistance 
for a blockchain and theoretically explain how tamper resistance is improved 
using our cross-referencing method. With our method, tamper resistance 
improves as the number of domains increases. 
 The proposed system of 1,000 domains are 3-10 times more tamper-resistant 
than that of 100 domains, and the capacity is 10 times higher. 
 We conclude that our method enables a more scalable and tamper-resistant 
public blockchain balanced with decentralization.
\end{abstract}


\end{frontmatter}

%\linenumbers

\section{Introduction}
\label{sec:intro}

 Since Bitcoin \cite{Bitcoin} appeared in 2008, blockchain technology 
has been gaining considerable public attention. 
 The words ``block chain'' were born from a dialogue between the 
Bitcoin inventor, Satoshi Nakamoto, and a cryptographer, Hal Finney, 
on the cryptography mailing list \cite{Finney2008}. 
 A blockchain is a chain of block-structured databases where each block 
is connected in a time-series order by a cryptographic hash function, 
and works as a timestamp server. 
A blockchain is not a new idea, as the citing of Massias 
\textit{et al.}'s paper published in 1999 \cite{MAQ1999} in Nakamoto's 
white paper suggests. However, it has become an iconic image for related 
technology. 

 All transactions in Bitcoin, including the issuance and exchange of 
electronic cash, are disclosed on the blockchain. 
 Therefore, nodes participating in the peer-to-peer (P2P) network can verify which 
address (not user) holds how many bitcoins by tracking transaction records. 
 However, since users can keep as many bitcoins addresses as they want, 
it is difficult for others to know how many bitcoins one owns in total. 
 Therefore, a minimum level of users' privacy is considered. 
 Personal transaction information has generally been kept private 
due to privacy concerns. 
 However, Bitcoin has made all transaction records public with privacy 
in mind (although anonymity is not guaranteed). 
 This makes it possible for an unspecified number of nodes to reach 
a (extended) consensus about transaction records on the blockchain 
on the basis of the longest chain rule. 
 Because this consensus algorithm was essentially a new idea, compared 
with known ones in distributed systems, 
it is called \textit{Nakamoto Consensus}.
% (仲基合意)

 Transaction processing is completed when a transaction is saved to 
the blockchain through a high-load computational process called 
proof of work (PoW) \cite{Back2002}. 
 Any node can become an \textit{authority} that connects a block with 
several transactions to the blockchain by presenting evidence to 
other nodes that it is able to execute the PoW correctly and quickly. 
 Therefore, the system is designed so that no particular node can remain 
as an unjustified authority. 
 In addition, PoW makes transaction records on the blockchain tamper-resistant. 

 The essential value of Bitcoin is \textit{micropayment}, which is possible 
by reducing the transaction fee to an extremely low level, 
such as less than one cent or one yen. 
 Micropayment has the potential to create a new decentralized economy, 
as it enables charging for various operations performed on the Internet. 
 For example, Wikipedia, which is always struggling to collect donations, 
may be able to collect server maintenance fees from very small 
donations from some article readers by introducing micropayment. 
 However, the current blockchain technology is practically incapable of 
supporting micropayment because the transaction processing capacity 
is restricted by the block transfer speed between nodes and the limit of 
block size, which is known as the \textit{Bitcoin scalability problem}. 

 Several technologies have been studied to solve this problem. 
 Off-chain scaling technologies, such as Lightning Network 
\cite{PD2016}, are currently attracting attention. 
 Off-chain technologies leave less transaction records in the blockchain. 
 Bitcoin has been used as a payment means of conducting 
illegal transactions in darknet markets, and there have 
been many reports on the managers and users of these markets being arrested 
\cite{silkroad,alphabay,welcome2video}.
 These arrests are attributable to Bitcoin's disclosure of all transaction 
records with tamper resistance, which are available as legal evidence. 
 If off-chain scaling technologies become widespread, 
transactions that cannot be audited by our societies and governments can be easily created and off-chain services might become 
hotbeds of illegal transactions in darknet markets for money laundering 
and financing of terrorism by criminal organizations. 
 Thus, when we consider the use of blockchain technology on the basis of 
law and ethics, it is ultimately necessary to solve the scalability problem 
on-chain.  
%--------------------------------------------------
 Existing blockchains, such as Bitcoin, are based on the premise that only
one blockchain can be globally integrated and managed. 
As a result, blocks are transferred and shared among nodes all over the world, 
which slows down transaction processing. 

As a way of speeding up transactions, we have proposed a mechanism for allocating
domains to geographically close nodes, similar to the country code top-level
domain in the Domain Name System, and managing blockchains in each domain
\cite{fujihara1,fujihara2,fujihara2019}. 
Through this mechanism. it is possible to distinguish between 
the inside and outside of a domain by the communication speed with the central node, 
which is the entrance to the P2P network.
 In addition, our mechanism allows domain-specific geographic information to be handled by connecting
secure Internet-of-Things devices to some of the nodes. 
However, the domain partition entailed by our mechanism reduces the number 
of nodes participating in each domain, which degrades the decentralization and tamper resistance of the blockchain.
 It is possible to distinguish between the inside and outside of a domain by 
the communication speed with the central node which is the entrance to the 
P2P network. 
 In addition, domain-specific geographic information can be handled by 
connecting secure Internet-of-Things devices to some of the nodes. 
 However, domain partition reduces the number of nodes participating in 
each domain, which degrades the decentralization and tamper-resistance 
of the blockchain. 

 To solve this problem, we previously proposed a cross-referencing method for 
periodically exchanging the state of the latest fixed block in the blockchain 
with hysteresis signatures \cite{susaki} among all the domains via the upper 
network layer \cite{EIDWT2021}. 
 We also designed a communication protocol to autonomously executes our 
cross-referencing method among domains. 
 We explained the effectiveness of this method only in terms 
of a definition of tamper resistance. 

 In this study, we evaluated the effectiveness of our 
method from the viewpoints of decentralization and scalability, as well as 
tamper resistance.
 We define decentralization as the ability of every participating node to 
affect the entire system by creating blocks.
 With our method, in-domain decentralization is equivalent to that of 
usual public blockchains, such as Bitcoin, while out-domain decentralization is 
preserved by hysteresis signatures, which affect some of the blocks 
in other domains. 
 Regarding scalability of transaction processing capacity, the performance of 
the entire system is improved because transactions and blocks are distributed 
only inside the domain. 
 We theoretically demonstrate that the transaction processing capacity should increase 
to that of a typical credit card system, such as 
VISA, meaning that the capacity can be evaluated by multiplying the number of 
domains by the average improvement ratio of transaction processing speed due to 
the increase in block size and reduction in the block-generation-time interval 
by domain partition. 
 For tamper resistance, each domain has evidence of the hysteresis signatures of 
the other domains in the blockchain. 
 Therefore, to tamper with a block in a domain, one must also tamper with 
the blocks containing the relevant hysteresis signatures in all domains. 
 We introduce two types of tamper-resistance-improvement ratios $R$ and $R'$
 and 
theoretically show that, in the terms of both measures, tamper resistance improves when using our cross-referencing 
method through Monte Carlo simulations. 
 We conclude that it is possible to achieve a more scalable and tamper-resistant 
public blockchain balanced with decentralization by using our cross-referencing method.

The main contributions of this paper are summarized as follows. 

\begin{itemize}
  \item We formulate the Bitcoin scalability problem to show an upper limit of 
	transaction processing capacity. 
  \item We clarify the definition of decentralization and elaborate that the 
	scalability of public blockchains potential to dramatically improved to increase 
	the transaction processing capacity to that of a VISA credit card system.
  \item We define two tamper-resistance-improvement ratios, $R$ and $R'$, to 
	explain that our cross-referencing method can also improve the tamper-resistance 
	of public blockchains. We also discuss how the ratios change when some 
	nodes fail.
\end{itemize}

\section{Related Work}
\label{sec:related}

 There are naive methods for solving the problem on-chain: increasing 
the block size and shortening the block-generation-time interval. 
 Both are possible by increasing the communication speed between 
nodes, but it is difficult for nodes with slow communication speed to join 
the network. 
 The former method is being evaluated on the Bitcoin Scaling Test Network 
(STN) \cite{bitcoinscaling}. 
 While the maximum block size of Bitcoin Core is 1 MB, STN has eliminated the block 
size limit and achieved an average processing speed of 2,596 transactions 
per second in 24 hours \cite{bitcoinscaling} (Accessed 13 April 2021). 
 The latter method is being evaluated by bloXroute \cite{bloX}, 
a project that proposes to alleviate the scalability problem by introducing a 
backbone network to propagate large blocks in a shorter time. 

 Hysteresis signatures \cite{susaki} enhance the tamper resistance of 
ordinary electronic signatures by adding a nested structure with other past 
electronic signatures. 
A typical structure of a hysteresis signature is shown in Fig. \ref{fig:hysteresis}. 
 A nested structure is naturally created by signing the content including 
the previous signature. 

 By repeating the nested structure, the electronic signatures are chained 
to create a time-series context. 
 For example, let the previous hysteresis signature be $S_{n-1}$ and 
the content to be signed be the block data of $m$ domains in total 
($B_{D_1}, \cdots, B_{D_m}$). 
 In this case, the hysteresis signature is created by signing the 
concatenation of the summary of the previous hysteresis signature $H(S_{n-1})$ 
and hashed contents $H(B_{D_1}), \cdots, H(B_{D_m})$. 
 The created signature $S_n$ is also added to the hysteresis 
signature, as shown in Fig.\ref{fig:hysteresis}.

Transaction signatures in Bitcoin was used 
to prove the legitimacy of a transaction by having the holder 
of the Bitcoin sign the transaction using his or her private key.
In hysteresis signature, the central core node of the domain signs 
the latest confirmed block in order to prove the validity of the block.
In addition, the summary finalized blocks of all past domains are aggregated and stored 
in the cross-referencing part, which enables tamper detection and correction among domains. 

 Ordinary electronic signatures can be tampered if the private key is leaked, 
and it is often impossible to detect tampering. 
 In hysteresis signatures, since the signature is signed 
in the nested structure, an attacker would have to tamper with all nested signatures 
after the tampered content, which makes tampering significantly more difficult. 
 This is similar to the block structure in the public blockchain. 
 If a contradiction in a hysteresis signature is found, 
it is almost certain that the content has been tampered with.

 BBc-1 \cite{saito} is a distributed ledger technology, with which 
a distributed system manages a consistent ledger, but it does not handle 
any public blockchain. 
 The system stores private transaction records in a tamper-resistant manner 
by using hysteresis signatures. 
 There is a reference implementation of the node, and the effectiveness 
of the system can be confirmed. 
 With BBc-1, all the transaction data have a hysteresis signature, which 
is exchanged across domains to increase tamper resistance. 
 The energy cost is also low compared with a typical public 
blockchain, such as Bitcoin, because of the absence of the PoW. 
 However, it is difficult to estimate how much tamper 
resistance is improved by introducing the hysteresis signature. 

 Atomic Swap is a technique that allows the exchange of cryptocurrencies 
recorded on different blockchains without needing to trust any third party 
\cite{atomicswap}. 
 This technique is also useful for coexisting multiple blockchains and 
exchanging their native cryptocurrencies.

\section{Cross-referencing Method}
\label{sec:teian}

\begin{figure*}[hbt!]
  \begin{center}
    \includegraphics[width=120mm]{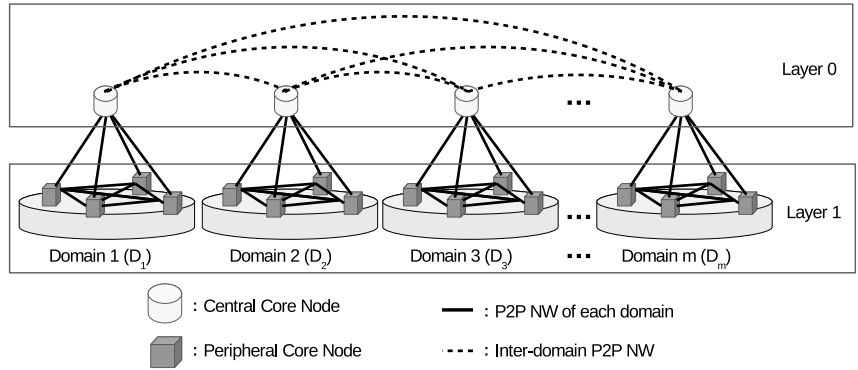}
  \end{center}
  \caption{Two-layer P2P network used in this study}
  \label{fig:p2p}
\end{figure*}

 The structure of the P2P network used in this study is shown in Fig. \ref{fig:p2p}. 
\begin{figure}[tb]
  \begin{center}
    \includegraphics[width=60mm]{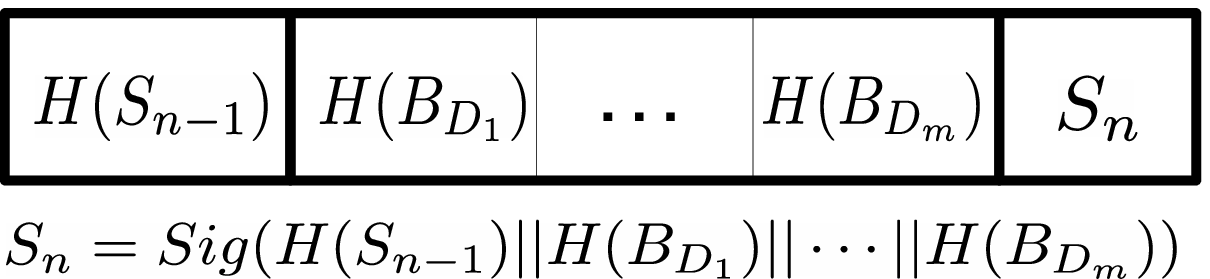}
  \end{center}
  \caption{Hysteresis signature}
  \label{fig:hysteresis}
\end{figure}
 The whole network consists of two P2P layers, \textit{i.e.}, Layers 0 and 1.
 Layer 1 assumes that there are multiple P2P networks of typical public blockchains which share 
transaction and block data. 
 Each P2P network in Layer 1 is called a \textit{domain}, and it is
assumed that there is a set of core nodes in each domain.
 There are two types of core node: central core nodes (CCNs) and peripheral 
core nodes (PCNs). 
 It is assumed that at least one CCN is selected as a leader in each domain 
beforehand.
 In this study, we assumed that the number of domains is $m$ and the CCNs 
of multiple domains ($D_1, D_2, \cdots, D_m$) have a prior agreement to 
share their block records and domain hysteresis signatures to use the 
cross-referencing method.
 In Layer 0, the CCNs are connected to each other to form another 
consortium-type P2P network, which is disconnected from that of Layer 1. 
 For simplicity, we consider the case in which the number of CCNs for each 
domain is one, but it is also possible to generalize the case in which 
the number of CCNs is more than one.
 Note that the addition of Layer 0 over Layer 1 is common with bloXroute 
\cite{bloX}, but the difference is that Layer 0 is also another P2P network 
with our method, while Layer 0 in bloXroute is a faster network 
transport layer for both transaction and block records.

\begin{figure}[htb!]
  \begin{center}
    \includegraphics[width=90mm]{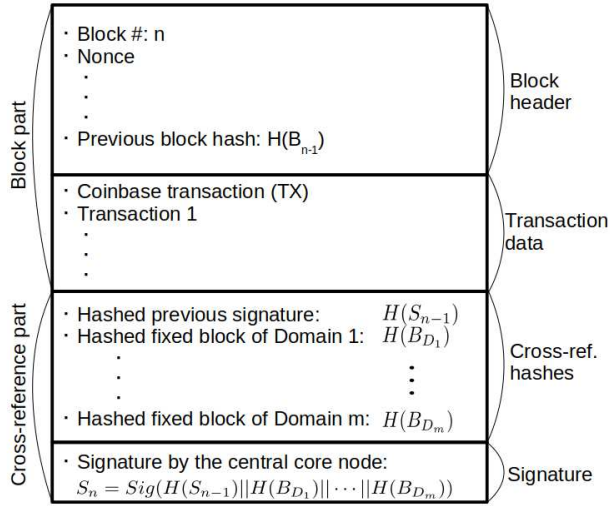}
  \end{center}
  \caption{Block structure of our cross-referencing method}
  \label{fig:block}
\end{figure}

 The block structure of our cross-referencing method is shown 
in Fig. \ref{fig:block}. 
 The difference between the block structure in ordinary public blockchains 
and that with our method is the cross-reference part. 
 In this part, a hysteresis signature is created by signing the concatenation 
of the summary of the previous hysteresis signature $H(S_{n-1})$ and 
hashed contents $H(B_{D_1}), \cdots, H(B_{D_m})$. 
 The created signature $S_n$ is the same as that shown in Fig.~\ref{fig:hysteresis}. 
The block records with the cross-reference part are shared between CCNs 
via the P2P network in Layer 0. 

\begin{figure}[htb]
  \begin{center}
    \includegraphics[width=90mm]{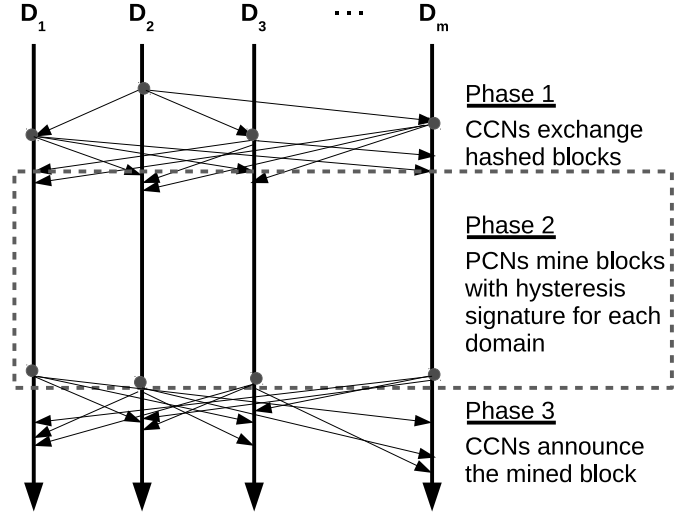}
  \end{center}
  \caption{Timeline of our cross-referencing method in normal case in which no stop 
	failure occurs in CCNs}
  \label{fig:cross-ref}
\end{figure}

 The timeline of our cross-referencing method in a normal case, 
meaning that no stop failure occurs in CCNs, is shown in Fig. \ref{fig:cross-ref}.
 The timeline is divided into three phases. 
 The details of each phase are explained as follows. 
\begin{enumerate}
\item In Phase 1, a CCN in a domain notifies the other CCNs to start 
      cross referencing by sending a message. 
      Then, each CCN transfers a message including the 
      $l$-confirmed block to the other CCNs, where $l$ is a positive integer 
      and $l$-confirmed means the block approved $l$ blocks before 
      the latest block. Phase 1 finishes if all the CCNs collect all the 
      $l-$confirmed block records by sharing them with each other. 

\item In Phase 2, each CCN first generates a hysteresis signature, as shown in 
      Fig. \ref{fig:hysteresis}. The CCN then sends a request message with 
      the hysteresis signature to PCNs in the same domain to mine the block 
      having the cross-reference component. 
      After independently mining the latest block, 
      a PCN sends a message with the mined block back to the CCN, and the CCN 
      checks whether the block has been properly mined. If the block has not been properly 
      mined, the CCN waits until a properly mined block is received and than Phase 2 finishes. 

\item In Phase 3, each CCN broadcasts the mined block to announce that 
      the cross-referencing was successful.
\end{enumerate}

\begin{figure}[hbt!]
  \begin{center}
    \includegraphics[width=80mm]{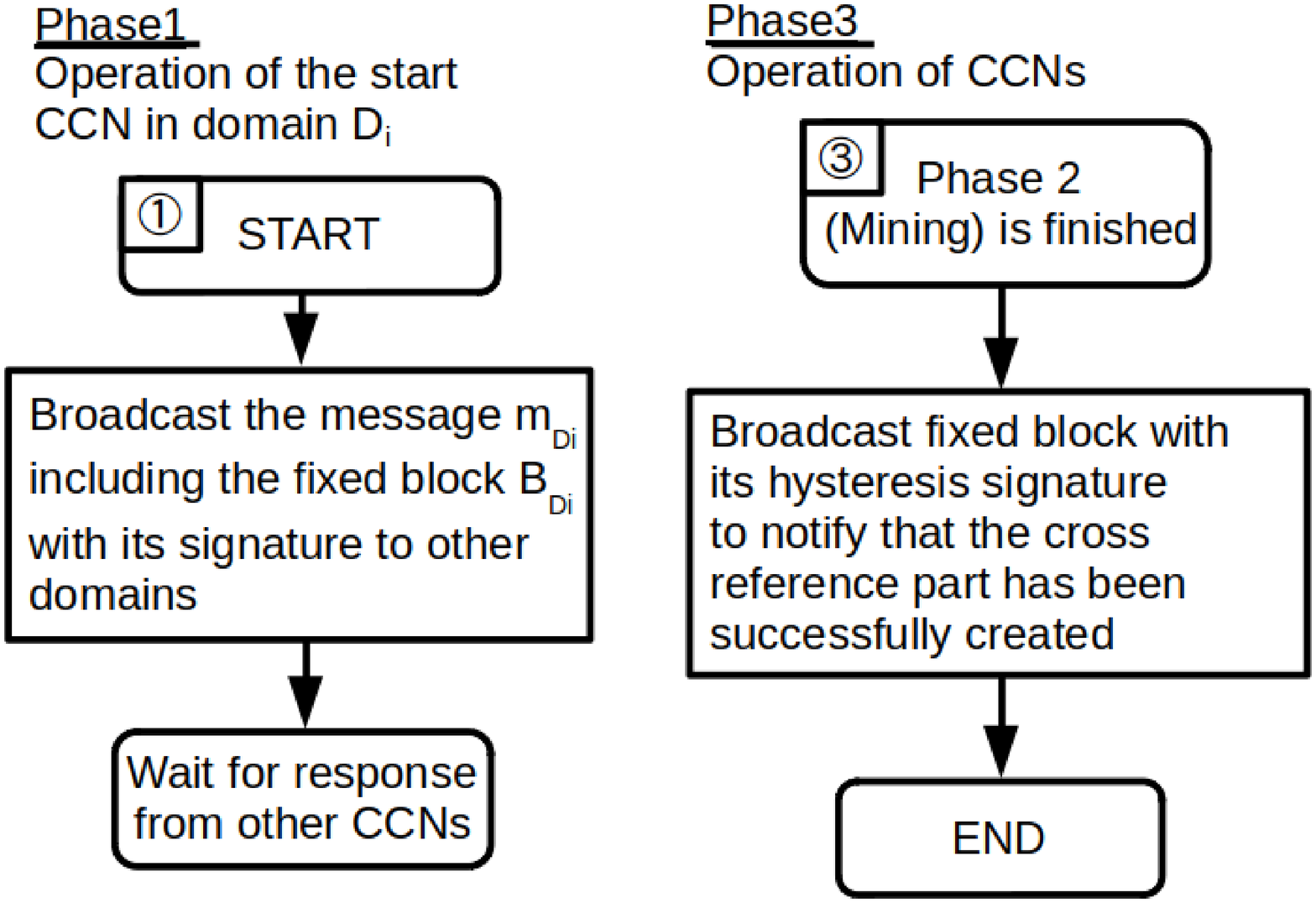}
    \includegraphics[width=80mm]{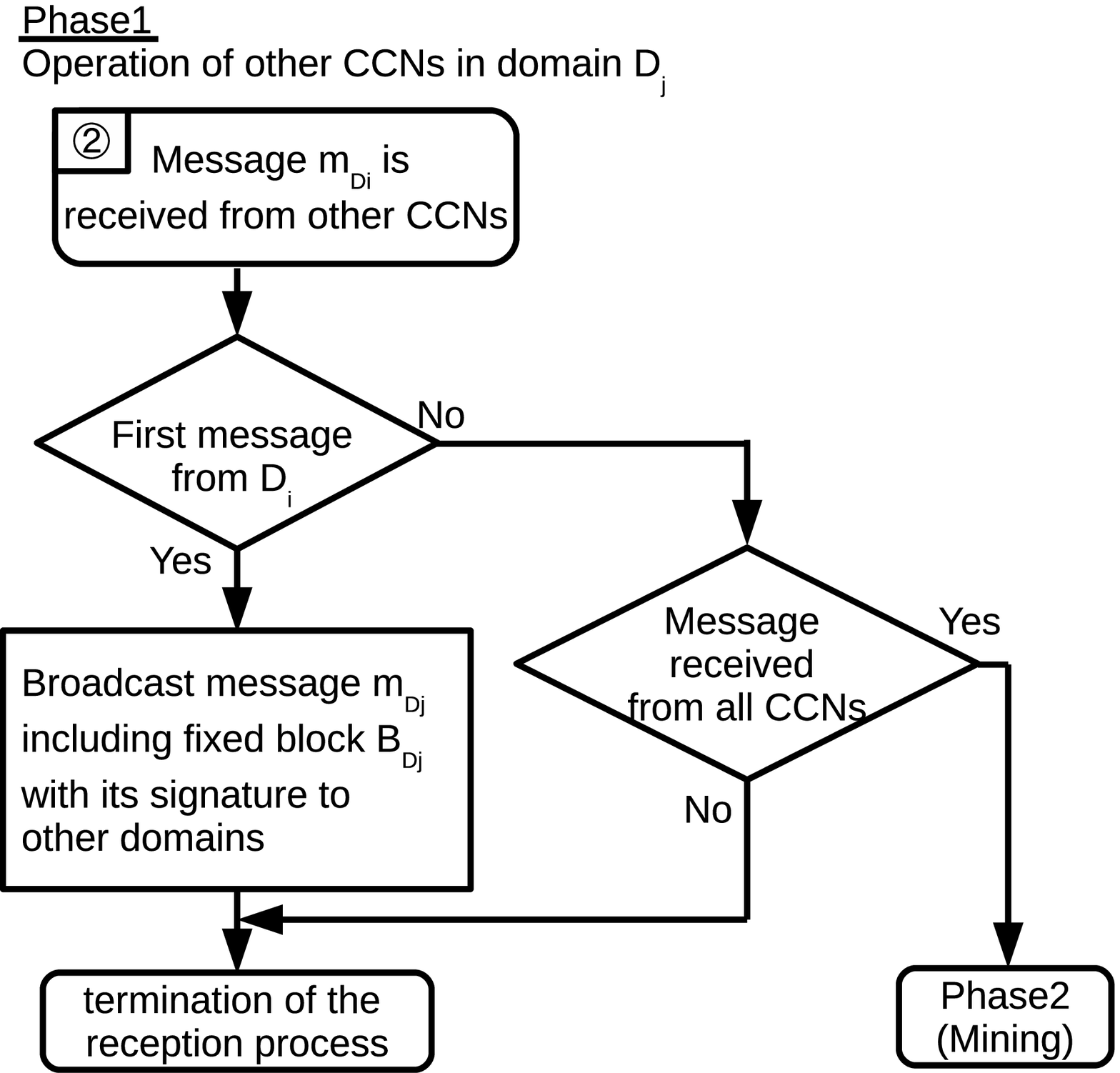}
  \end{center}
  \caption{Flowchart 1 (Phase 2 is omitted because it consists 
	of usual mining process on Layer 1)}
  \label{fig:algorithm1}
\end{figure}

\begin{figure}[hbt!]
  \begin{center}
    \includegraphics[width=90mm]{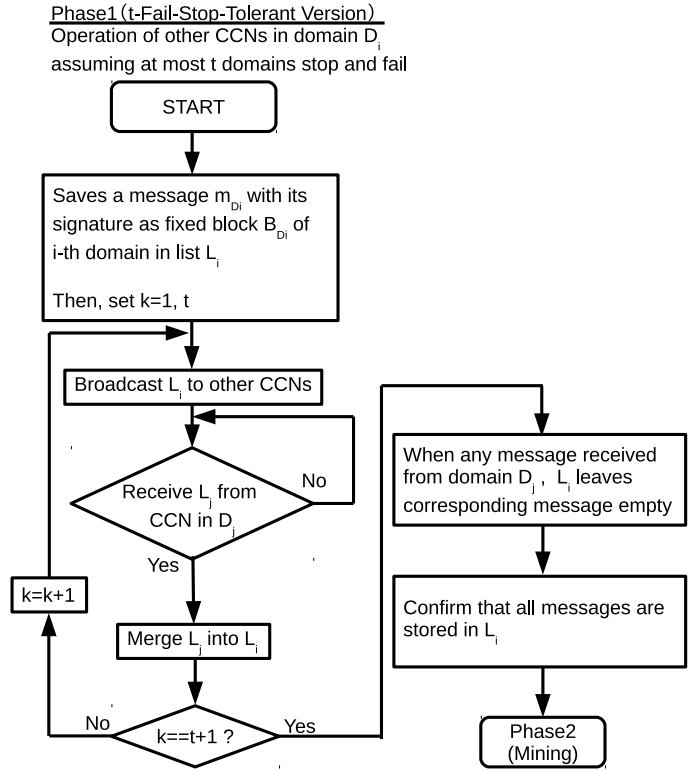}
  \end{center}
  \caption{Flowchart 2 (Phase 1 only; other phases are common to those 
	   in Flowchart 1)}
  \label{fig:algorithm2}
\end{figure}

 We designed a distributed algorithm \cite{KS2011} for our cross-referencing 
method. The details of which are shown in Fig.~\ref{fig:algorithm1}.
 We also considered a distributed algorithm that is tolerant of $t$-stop 
failures (CCNs in $t$ domains do not respond because they experienced 
a stop failure or refuse to execute cross referencing) as shown in 
Fig.~\ref{fig:algorithm2}.

 There are three assumptions under which these flowchart work properly. 
\begin{enumerate}
\item The P2P network in Layer 0 is synchronous and its structure should 
      be a complete graph. 
\item All the CCNs are reliable, meaning that they execute the cross-referencing 
      method following the flowchart to share the requested block data with 
      each other.
\item In Flowchart 1, none of the CCNs exhibit any experience stop failure and in Flowchart 2, 
      The CCNs allow $t$-stop failures, which means that our cross-referencing method 
      works even if at most $t$ CCNs do not cooperate to share their block records. 
\end{enumerate}
 Note that we can assume that the number of nodes having experiencing stop failure is small because 
each domain needs to strengthen tamper resistance with 
the cross-referencing method. 
 If it takes a long time to repair a failed CCN, the blockchain in its 
domain becomes vulnerable to malicious tampering attacks. 
 Therefore, each domain should recover from the stop failure as soon as possible. 

 The efficiency of these distributed algorithms can be evaluated by measuring 
both communication and time complexity. 
 The communication complexity of Flowchart 1 is $O(m^2 \cdot b)$, where $m$ is 
the number of CCNs (equivalent to the number of domains), and $b$ is the
bit size of the message to send between CCNs.
 The communication complexity of Flowchart 2 is $O((t+1)(m^2 \cdot b))$.
Assume that the time complexity of Flowchart 1 is $ T_1 + T_2 + T_3 $, where 
$T_i (i=1, 2, 3)$ is the waiting time taken for Phase $i$ in Flowchart 1. 
Then, the time complexity of Flowchart 2 is $(t+1)T_1 + T_2 + T_3 $.

\section{Theoretical Evaluation}
\label{sec:theory}%
 Our cross-referencing method can improve the scalability and tamper resistance of 
the blockchain while preventing degradation of decentralization. 
 We theoretically evaluated the effectiveness of the proposed method from 
three viewpoints: decentralization, scalability, and tamper resistance.

 \subsection{Evaluation of decentralization}
\label{sec:theory1}
 The definition of \textit{decentralization} is often ambiguous, 
so, we will define it as the ability of every participating node 
to affect the entire system by creating blocks.
 With our method, in-domain decentralization is equivalent to that of usual 
public blockchains, such as Bitcoin. 
 Out-domain decentralization is preserved by hysteresis signatures, which affect 
some of the blocks in other domains. 
 Note that this governance structure between in-domain and out-domain 
decentralization is similar to modern democratic systems in our society.

\subsection{Theoretical formulation of Bitcoin scalability problem} 

The block-generation-time interval of blockchain $T$ is a stochastic variable, 
and $T$ obeys the exponential distribution, \textit{i.e.}, 
\begin{equation}
  F(t) = P(T \le t) = \int_0^t \lambda e^{-\lambda t^{\prime}} dt^{\prime} = 1 - e^{-\lambda t},  \label{fork1}
\end{equation}
where $\lambda$ is the inverse of the average block-generation-time interval
\cite{DW2013}. 
 In Bitcoin Core, the average interval $1/\lambda$ is $10$ minutes 
($600$ seconds). 
The latency required to transfer a block of size $b = 1$ MB in Bitcoin 
Core to 90\% of the nodes on the P2P network was experimentally evaluated as 
$t = \tau_{fork}(b) \fallingdotseq 12$ seconds \cite{bloX}. 
 Therefore, the probability of a blockchain fork (branching) is estimated as 
\begin{eqnarray}
  F( \tau _{fork} ) = P(\tau_{fork}) = 1 - e^{-\lambda \tau _{fork}} 
	\fallingdotseq \lambda \tau _{fork} = 12/600 = 0.02. \label{fork2}
\end{eqnarray}
 If the blockchain forks, some of the block-generation capacities are wasted 
because one of the branches will be rejected. 
 Therefore, the fork probability is kept small enough by using 
a difficulty adjustment algorithm \cite{NOH2019}. 

 There are two simple ways of increasing the transaction processing capacity: 
(1) increasing the block size $b$ and (2) shortening the average block-generation-time interval 
$1/\lambda$. 
 As shown in Eq.~(\ref{fork2}), however, the fork probability increase with 
both methods, which means the transaction processing capacity is reduced. 
 Therefore, it is generally difficult to resolve this tradeoff of transaction 
processing capacity, which is called the scalability problem in 
public blockchains. 

 It is estimated that the transaction processing capacity of Bitcoin is 5-7 
transactions per second. 
 However, it is known that the credit card company VISA, Inc. has a capacity of 
56,000 transactions per second. 
 We explain below that it is difficult for Bitcoin Core to reach this level 
of the capacity. 
 From Eq.~(\ref{fork2}), the unfork probability, which only contributes to 
creating valid blocks, is calculated as 
\begin{equation}
  P_{unfork} = 1 - P(\tau_{fork}) = e^{-\lambda \tau_{fork}} = e^{-\tau_{fork}/\tau}, 
\end{equation}
where $\tau = 1/\lambda$ is the average block-generation-time interval. 
 In this case, the blockchain does not fork and it is effective on the 
transaction processing capacity. 
 Therefore, the transaction processing capacity per second is roughly  
\begin{equation}
  G(\tau) = \frac{C}{\tau} e^{-\tau_{fork}/\tau}, \label{eq:capacity1}
\end{equation}
where $C$ is the average number of transactions in a block. 
 In Bitcoin Core, $\tau=600$ seconds, $\tau_{fork} = 12$ seconds, 
the unfork probability is $e^{-\tau_{fork}/\tau} = 1 - 0.02 = 0.98$, and 
$G(\tau=600) = 7$ transactions per second at maximum. 
 Substituting these values into Eq.~\ref{eq:capacity1}, the average number 
of transactions in a block is about $C=4,286$ transactions. 

 We can further consider the optimal transaction processing 
capacity $\tau_{opt}$. 
 To this end, we estimate the maximum $G(\tau)$ by changing $\tau$, 
\textit{i.e.}, 
\begin{equation}
  \max_{\tau} G(\tau) = C \max_{\tau}( e^{-\tau_{fork}/\tau} / \tau ). \label{eq:capacity2}
\end{equation}
 It is easily calculated that $\tau_{opt} = \tau_{fork} = 12$ 
seconds. 
Therefore, the optimal transaction processing capacity per second is calculated as 
\begin{equation}
  \max_{\tau} G(\tau) = G(\tau_{fork}) = C e^{-1} / \tau_{fork}. \label{eq:capacity2}
\end{equation}
 Substituting $C = 4286$ transactions and $\tau_{fork} = 12$ seconds into 
Eq.~(\ref{eq:capacity2}), 
the optimal $G$ turns out to be about 132 transactions per second, 
which is considered to be the upper bound of the transaction processing capacity 
in the Bitcoin Core blockchain. 

\subsection{Improvement in scalability of transaction processing capacity}
\label{sec:theory2}

 As we explained in the previous subsection, an upper bound to the transaction 
processing capacity seems to exist, which causes the scalability problem. 
 Existing blockchains including Bitcoin are based on the premise that only 
one chain is globally integrated and managed. 
 Therefore, it is necessary to transfer and share blocks among nodes 
located all over the world, which dramatically reduces the transaction 
processing capacity. 
 Therefore, we considered a mechanism for allocating domains to geographically 
close nodes and managing one chain for each domain. 

 Regarding scalability in transaction processing speed, the performance of 
the entire system can be improved because transactions and blocks are 
distributed only inside the domain. 
 For example, if the block size is increased from 1 to 10 MB, block-generation-time 
interval from 10 to 2 minutes, and number of 
domains from 1 to 200, the transaction processing capacity can reach an
equal or better level compared with that of VISA, Ltd.
\begin{eqnarray}
  5 \textrm{tx./sec.} &\times& (\frac{10}{1}) \times (\frac{10}{2}) \times (\frac{200}{1}) = 50,000 \textrm{tx./sec.} \simeq 56,000 \textrm{tx./sec.} \label{BTCeq1} \\
  7 \textrm{tx./sec.} &\times& (\frac{10}{1}) \times (\frac{10}{2}) \times (\frac{200}{1}) = 70,000 \textrm{tx./sec.} > 56,000 \textrm{tx./sec.} \label{BTCeq2}
\end{eqnarray}
It is highly possible that the transaction processing capacity will exceed that of VISA credit card system, 
but the actual performance needs to be evaluated experimentally, which is left for our future work.

\subsection{Evaluation of tamper resistance} \label{sec:theory3}

 Our cross-referencing method can improve the tamper resistance of blockchains. 
 We will use two tamper-resistance-improvement ratios $R$ and $R'$ to 
evaluate tamper resistance. 
 Let the total number of core nodes in the distributed system be $N$ and the hash 
rate of core node $i$ be $h_i$. 
 Note that the hash rate is the number of times that a cryptographic hash function 
can be computed per unit time. 

 With $R$, tamper resistance can be estimated by the maximum hash rate 
because we assume that the node with the highest hash rate has the highest 
probability of generating blocks, so it continues mining. 
 On the other hand, most nodes with relatively smaller hash rates have a smaller 
probability of generating blocks, so they tend to quit mining. 
 Therefore, tamper resistance is proportional to the following value, 
\textit{i.e.}, 
\begin{equation}
	\max\{h_1, h_2, \cdots, h_N\}.
\end{equation}

 Suppose that the total number of core nodes in the $m$-th domain is $D_m$. 
 The tamper resistance of each domain is given by the maximum hash power of 
the nodes for each domain, \textit{i.e.}, 
\begin{eqnarray}
	A_1 &=& \max\{h_{11}, \cdots, h_{1D_1}\}, \\
	A_2 &=& \max\{h_{21}, \cdots, h_{2D_2}\}, \\
	    & & \vdots \nonumber \\
	A_m &=& \max\{h_{m1}, \cdots, h_{mD_m}\}. 
\end{eqnarray}
 When only one domain has $N=D_1+D_2+\cdots+D_m$ core nodes, 
the tamper resistance of the domain without the cross-referencing method 
can be estimated as 
\begin{equation}
  A = \max\{A_1, A_2, \cdots, A_m\}. \label{equation8}
\end{equation}

 With $R'$, tamper resistance can be estimated by the accumulated 
hash rates of the top $X$\% nodes with high hash rates. 
 Tamper resistance is then proportional to the following value, 
\textit{i.e.}, 
\begin{equation}
  A' = \sum_{i~\in~nodes~with~top~X\%~hash~rate} h_i. \label{equation9}
\end{equation}
 Equation (\ref{equation8}) evaluates the maximum hash rate as tamper resistance, 
and Eq. (\ref{equation9}) is an expression that evaluates the sum of the hash 
rates of the top $X$\% as tamper resistance.

 By applying our cross-referencing method among $m$ domains, 
tamper resistance can be estimated by the sum of all the highest hash rates 
in the domains, \textit{i.e.}, 
\begin{equation}
    \label{equation10}
	B = \sum_{i=1}^{m} A_i, 
\end{equation}
\begin{equation}
    \label{equation11}
	B' = \sum_{i=1}^{m} \sum_{j~\in~nodes~with~top~X\%~hash~rate} A_{ij},
\end{equation}

 Similarly, Eq. (\ref{equation10}) is an expression that evaluates the maximum 
hash rate in the domain as tamper resistance.
 Equation \ref{equation11} evaluates the sum of the hash rates of the top $X$\% 
that contributes to tamper resistance.

 Therefore, $R$ is defined as
\begin{equation}
    \label{R}
  R = \frac{B}{A} (> 1), 
\end{equation}
 and, $R'$ is defined as 
\begin{equation}
    \label{R'}
  R' = \frac{B'}{A'}. 
\end{equation}

 To estimate typical values of $R$ and $R'$, we conducted Monte Carlo simulations in which 
the hash rate $ h_{ij} $ ($i$ is the domain number, and $j$ is the serial number 
of nodes in a domain) was randomly assigned in accordance with a Pareto distribution, 
\textit{i.e.}, 
\begin{equation}
    \label{P}
	P(h_{ij}) = \frac{\alpha}{h_{ij}^{1+\alpha}} (h_{ij} > 1),
\end{equation}
where $\alpha$ is a scale parameter. 
 This distribution is often used to explain wealth distribution in economics. 
 The rate represents the total amount of computational resources, which is 
proportional to the amount of capital of the miner, so using the Pareto 
distribution is considered appropriate. 

It is known that the tamper resistance of  a blockchain for 
PoW is determined by the hash rate of the mining nodes on the P2P network 
\cite{hashrate}. 
 This hash rate is generally considered to follow the Pareto distribution 
in Eq. (\ref{P}) 
because computing resources of nodes are unequal. 
 Therefore, the hash rate depends on the financial strength of the miner who
owns the nodes. Also, since only a handful of miners succeed in mining on a regular basis, we defined 
the tamper resistance improvement ratio as shown in Eqs. (\ref{R}) and (\ref{R'}).

 We assume that the total number of core nodes is $N=10,000$ and that the number 
of core nodes in each domain is uniform, \textit{i.e.}, 10, 100, and 1000 nodes 
when the number of domains is 1000, 100, and 10, respectively.
 Typical simulation results are shown below.
%in Figs.~\ref{fig:alpha2}, \ref{fig:alpha3}
%\ref{fig:alpha2-top10}, \ref{fig:alpha3-top10}, \ref{fig:alpha2-top30} 
%and \ref{fig:alpha3-top30}. 

\begin{figure}[p]
  \vspace{-35mm}
  \begin{center}
    \includegraphics[width=80mm]{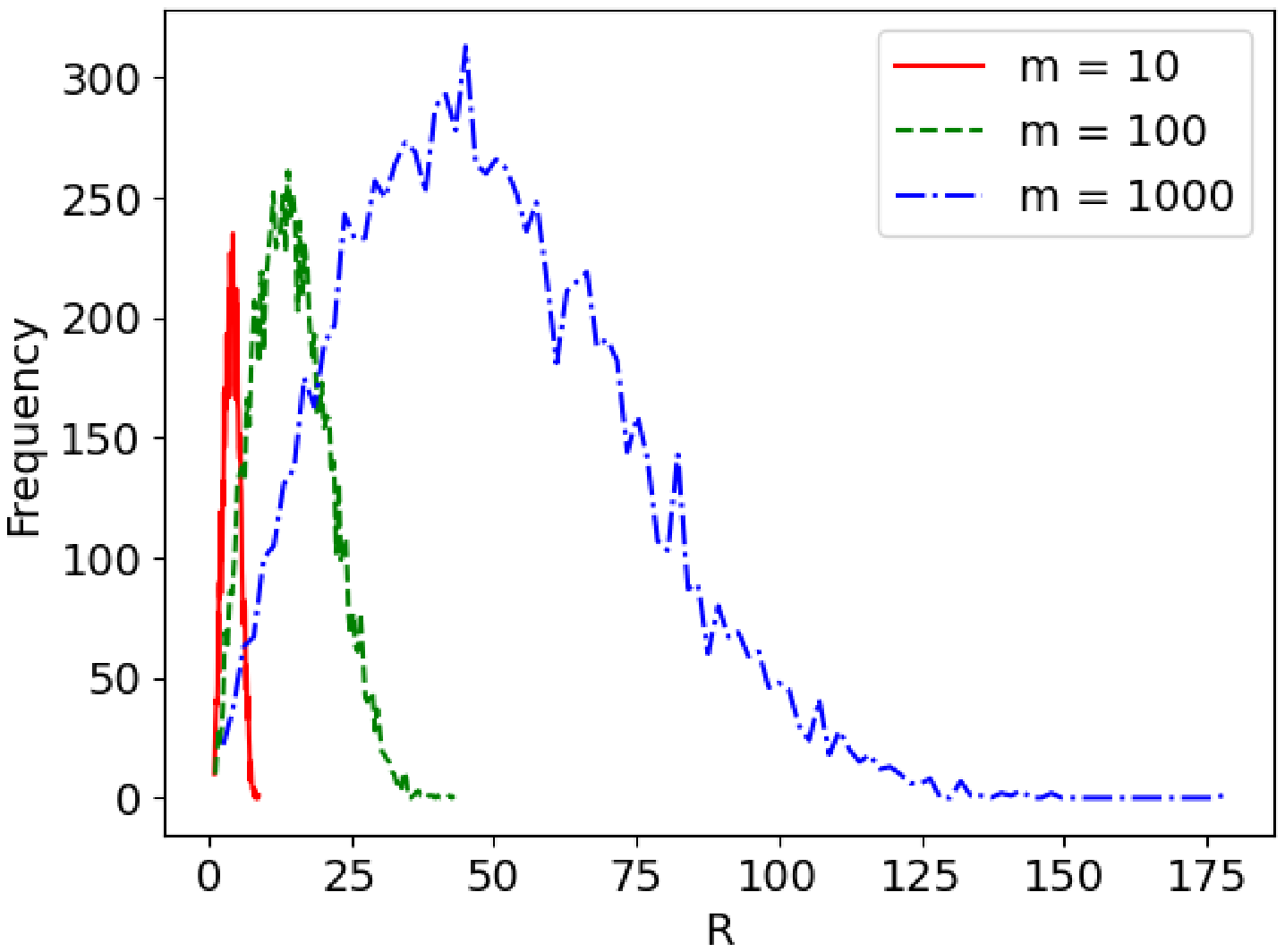}
  \end{center}
  %\vspace{35mm}
  \caption{Simulation results of probability density function of 
	tamper-resistance-improvement ratio $R$ when scale parameter $\alpha=2$}
  \label{fig:alpha2}
  \vspace{10mm}
  \begin{center}
    \includegraphics[width=80mm]{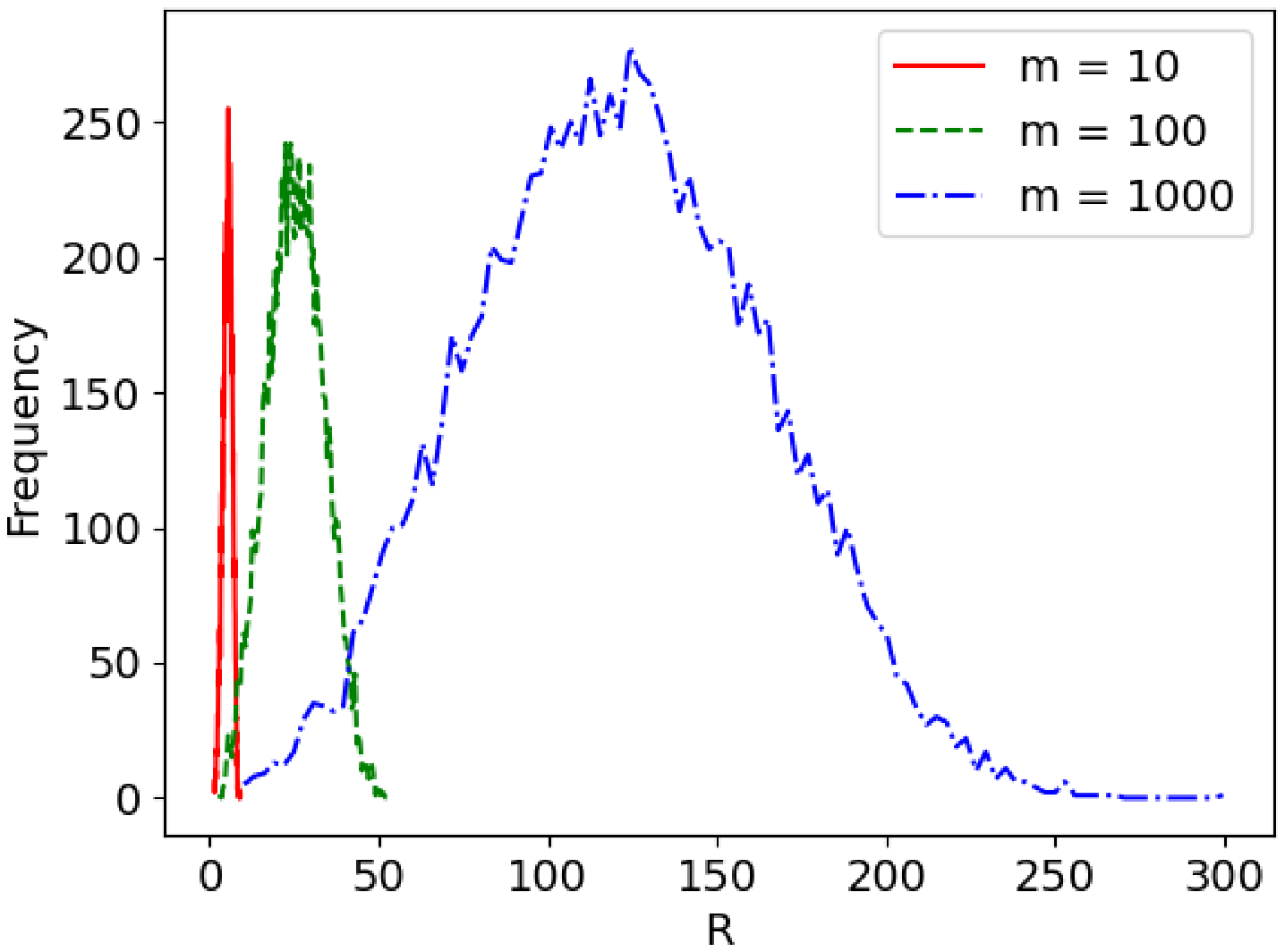}
  \end{center}
  %\vspace{35mm}
  \caption{Simulation results of probability density function of $R$ when $\alpha=3$}
  \label{fig:alpha3}
\end{figure}

\begin{figure}[p]
  %\hspace{-35mm}
  \vspace{-35mm}
  \begin{center}
    \includegraphics[width=80mm]{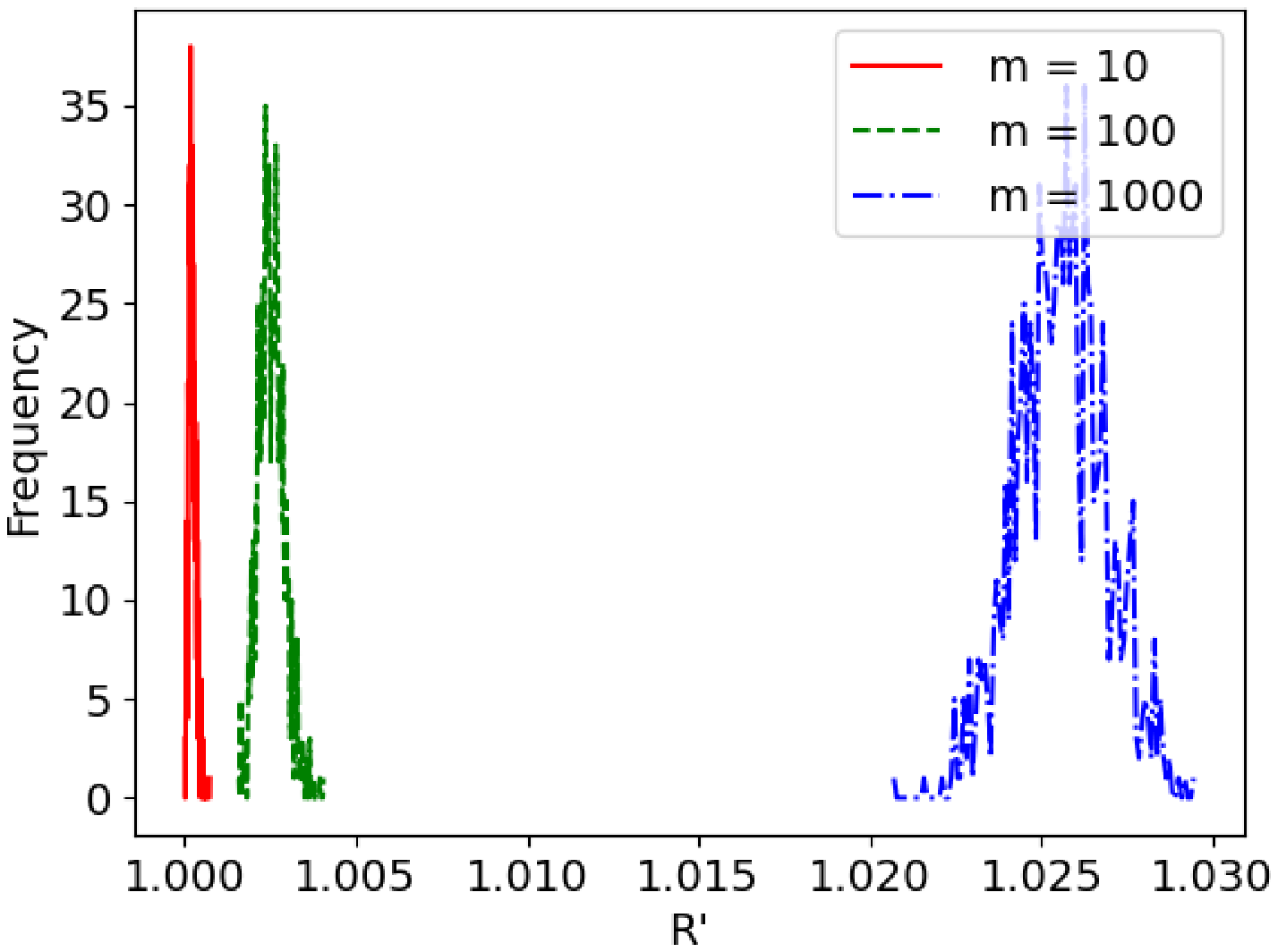}
  \end{center}
  %\vspace{35mm}
  \caption{Simulation results of probability density function of $R'$ when $\alpha=2$ and $X=10$\%}
  \label{fig:alpha2-top10}

  \vspace{10mm}
  \begin{center}
    \includegraphics[width=80mm]{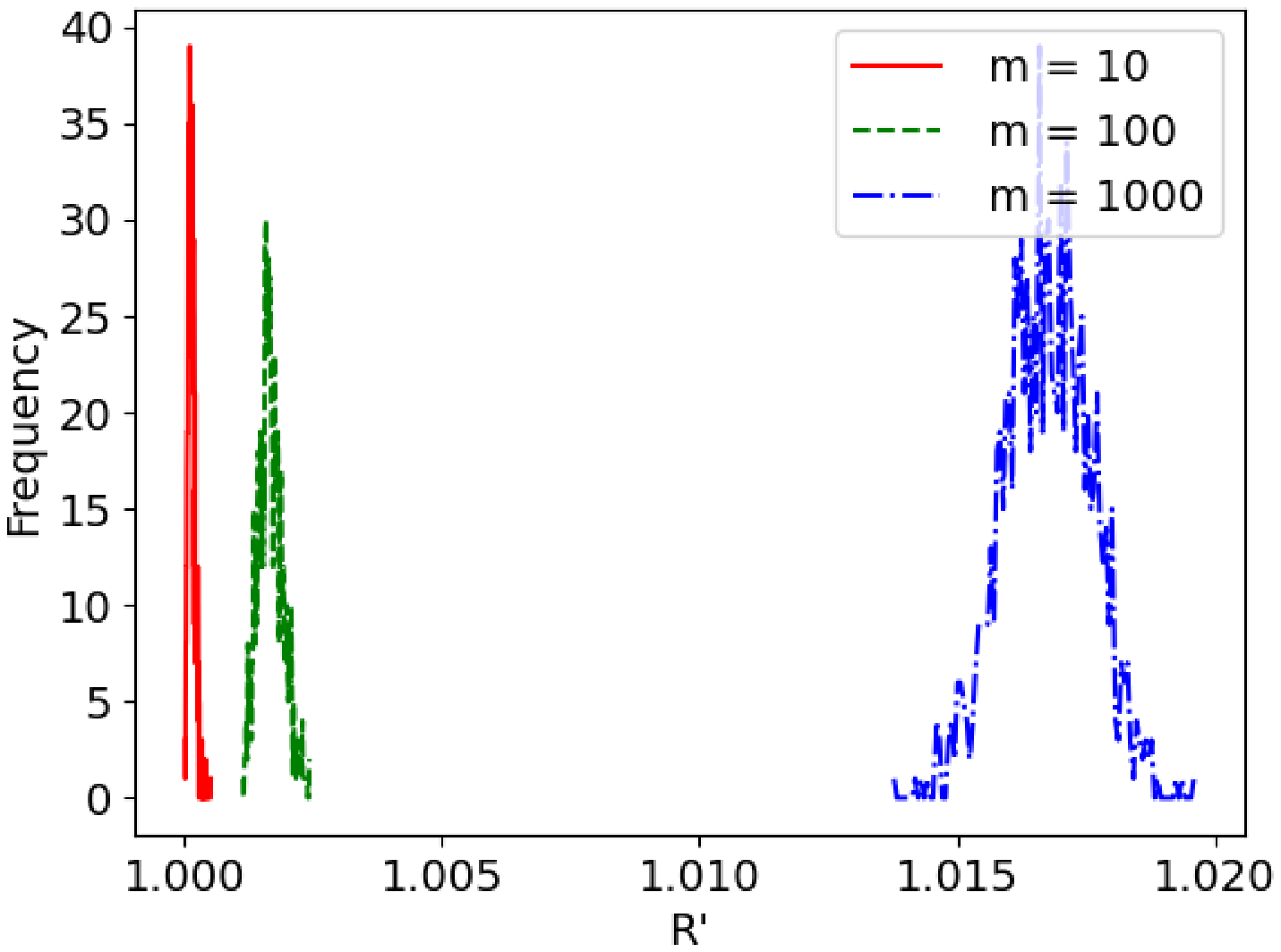}
  \end{center}
  %\vspace{35mm}
  \caption{Simulation results of probability density function of $R'$ when $\alpha=3$ and $X=10$\%}
  \label{fig:alpha3-top10}
\end{figure}

\begin{figure}[p]
  \vspace{-35mm}
  \begin{center}
    \includegraphics[width=80mm]{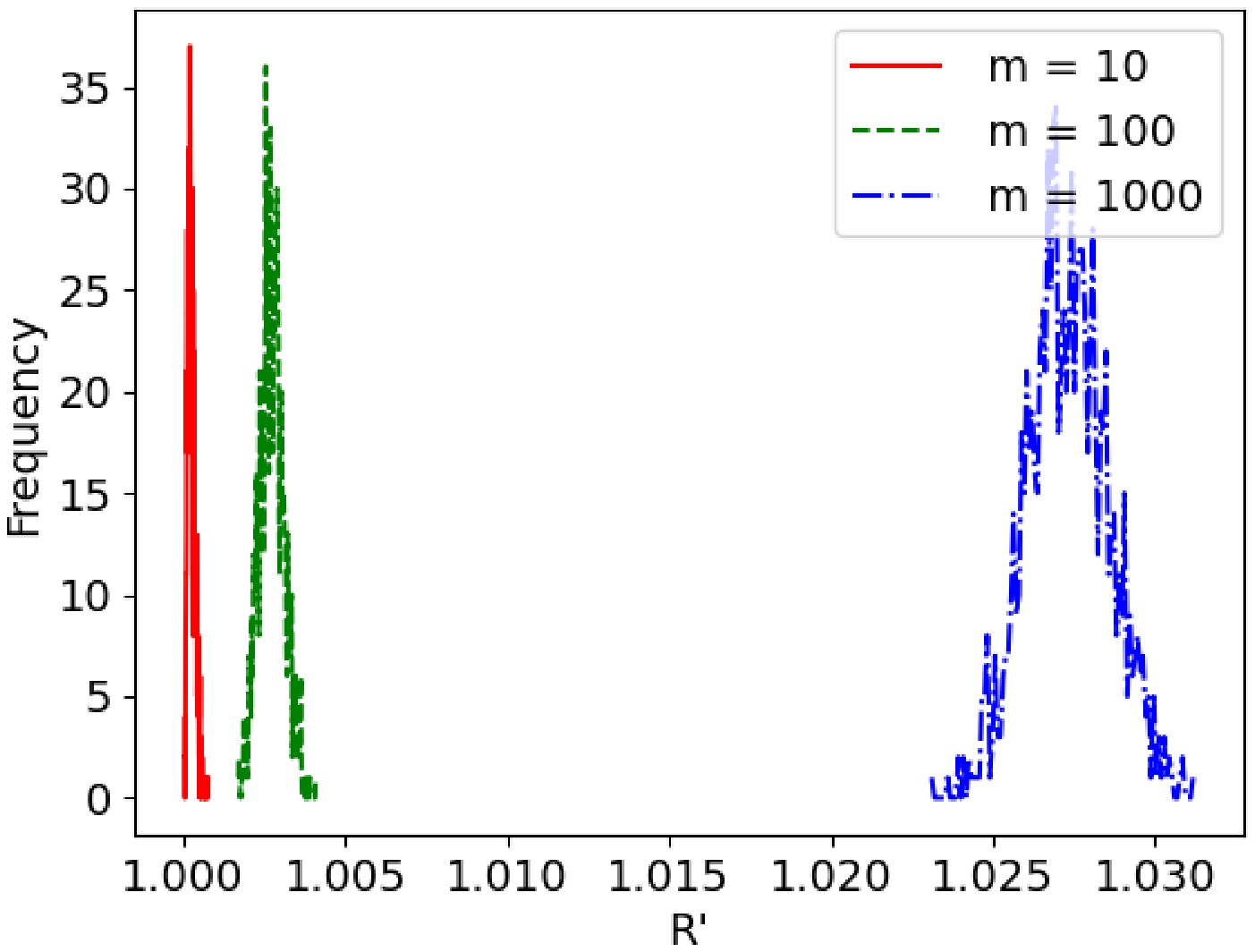}
  \end{center}
  %\vspace{35mm}
  \caption{Simulation results of probability density function of $R'$ when $\alpha=2$ and $X=30$\%}
  \label{fig:alpha2-top30}
  \vspace{10mm}
  \begin{center}
    \includegraphics[width=80mm]{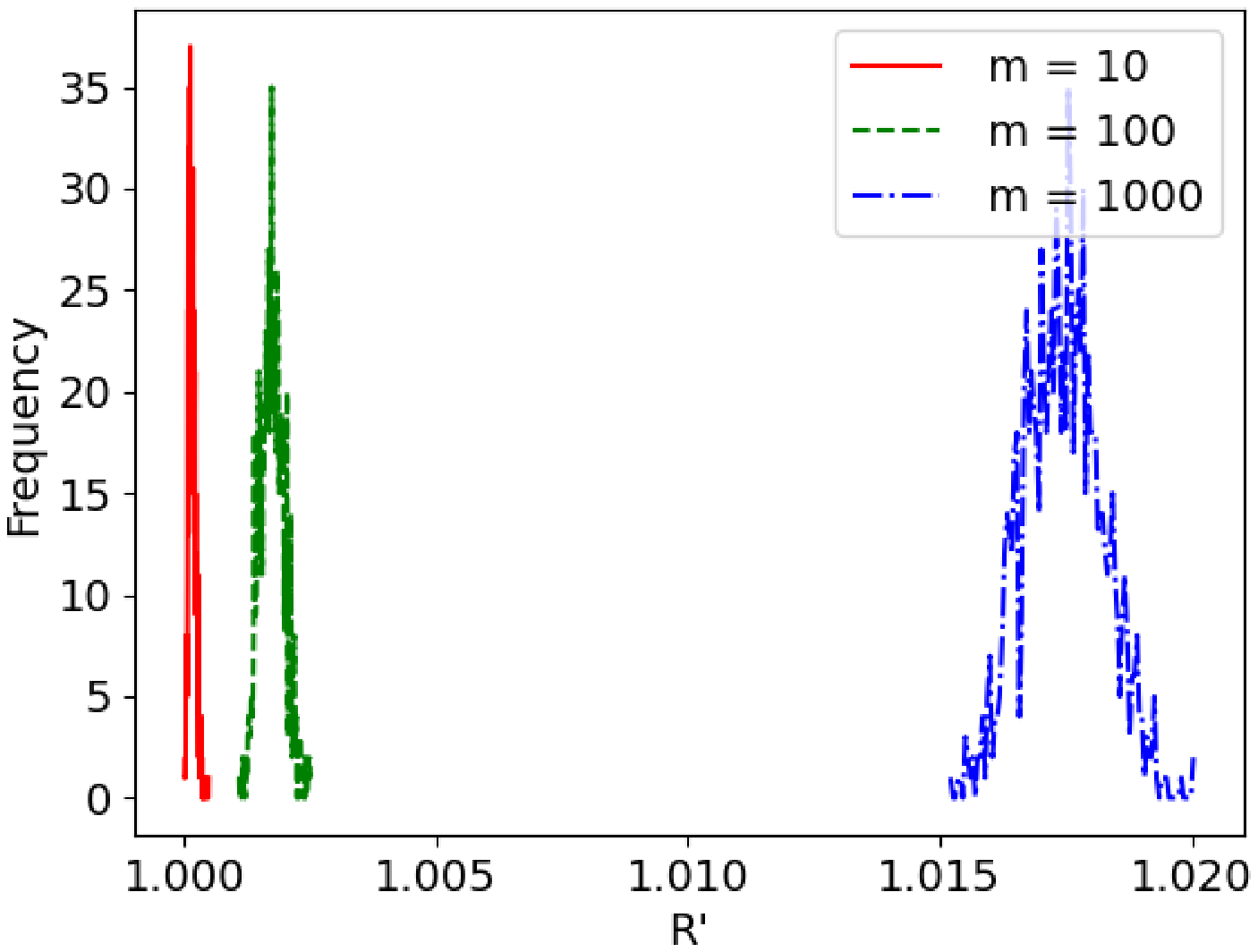}
  \end{center}
  %\vspace{35mm}
  \caption{Simulation results of probability density function of $R'$ when $\alpha=3$ and $X=30$\%}
  \label{fig:alpha3-top30}
\end{figure}

 Fig. \ref{fig:alpha2} and \ref{fig:alpha3} show the results when 
the maximum hash rate is used as tamper resistance. 
 Fig. \ref{fig:alpha2-top10} and \ref{fig:alpha3-top10} show the results of
$R'$ with the top $X=10$\%. 
 Fig. \ref{fig:alpha2-top30} and \ref{fig:alpha3-top30} show the results of
$R'$ with the top $X=30$\%. 
 A comparison of Figs.~\ref{fig:alpha2-top10}, \ref{fig:alpha3-top10}, 
\ref{fig:alpha2-top30} and \ref{fig:alpha3-top30} indicates that 
the peak of the distribution of $R'$ is the position slightly higher than one. 
 As $m$ increases, the peak of the histograms 
in both $R$ and $R'$ generally shifts to a higher position. 
As m increases, the tamper resistance gets higher.
In Figs. \ref{fig:alpha2} and \ref{fig:alpha3}, the tamper resistance with m=1,000 is about 3-5
times higher than that with m=100. In Figs. \ref{fig:alpha2-top10} and \ref{fig:alpha2-top30}, comparing
the peak positions of the distribution of R', we can see that the tamper
resistance with m=1,000 is 10 times higher that that with m=100.
In Figs. \ref{fig:alpha2-top10} and \ref{fig:alpha2-top30}, there is no significant difference between them in
both the top 10\% and 30\%. There is similar tendency in Figs. \ref{fig:alpha3-top10} and \ref{fig:alpha3-top30}.
In addition, these figure show that the variance of the distribution increases as 
$m$ increases. 

\newpage
\subsection{Effect of stop failures on tamper resistance} \label{sec:theory4}

\begin{figure}[p]
  \vspace{-35mm}
  \begin{center}
    \includegraphics[width=80mm]{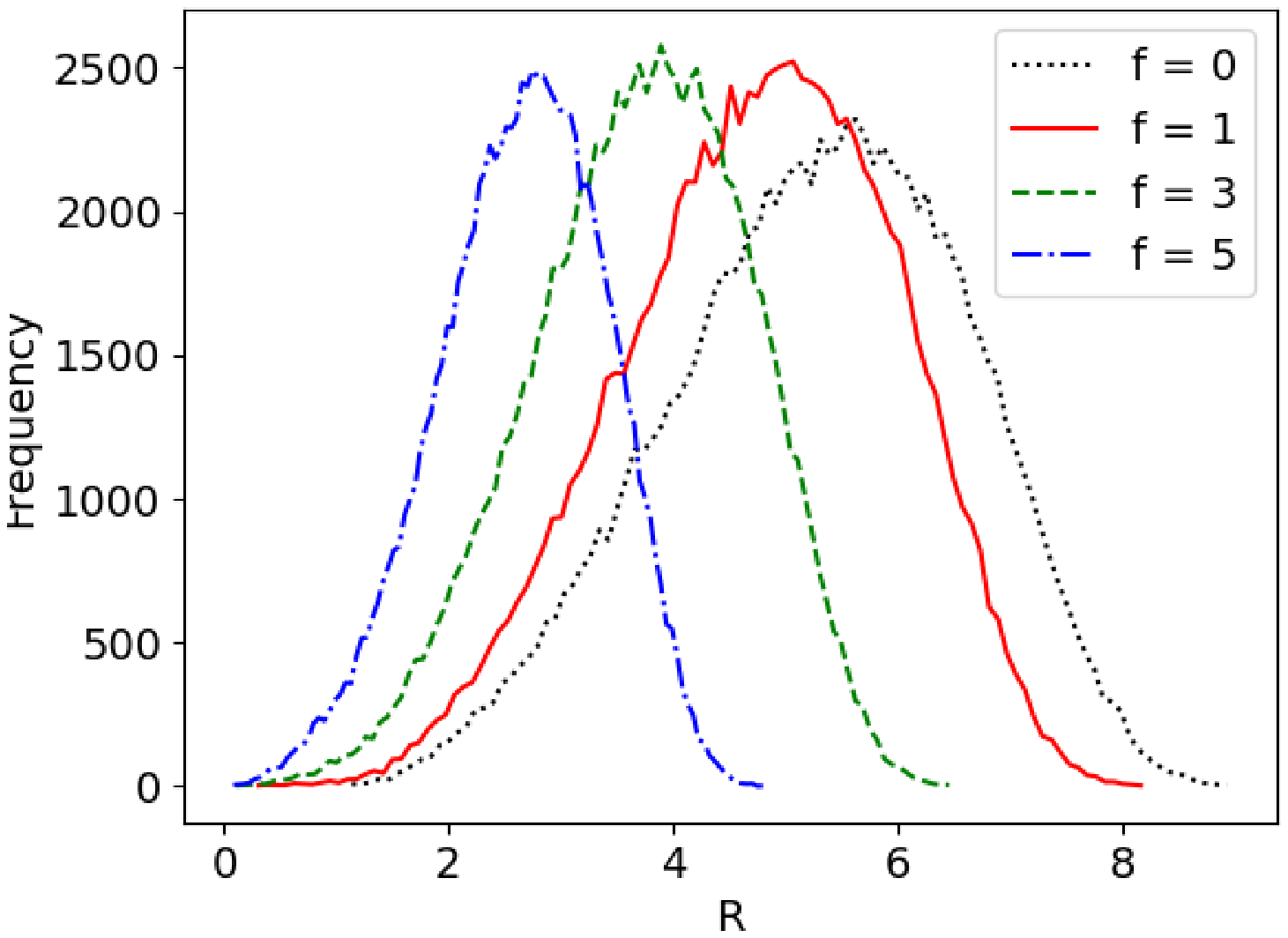}
  \end{center}
  %\vspace{35mm}
  \caption{Simulation results of probability density function of $R$ when $m=10$ and $\alpha=2$}
  \label{fig:alpha2-domain10-stopfailure}
  \vspace{10mm}
  \begin{center}
    \includegraphics[width=80mm]{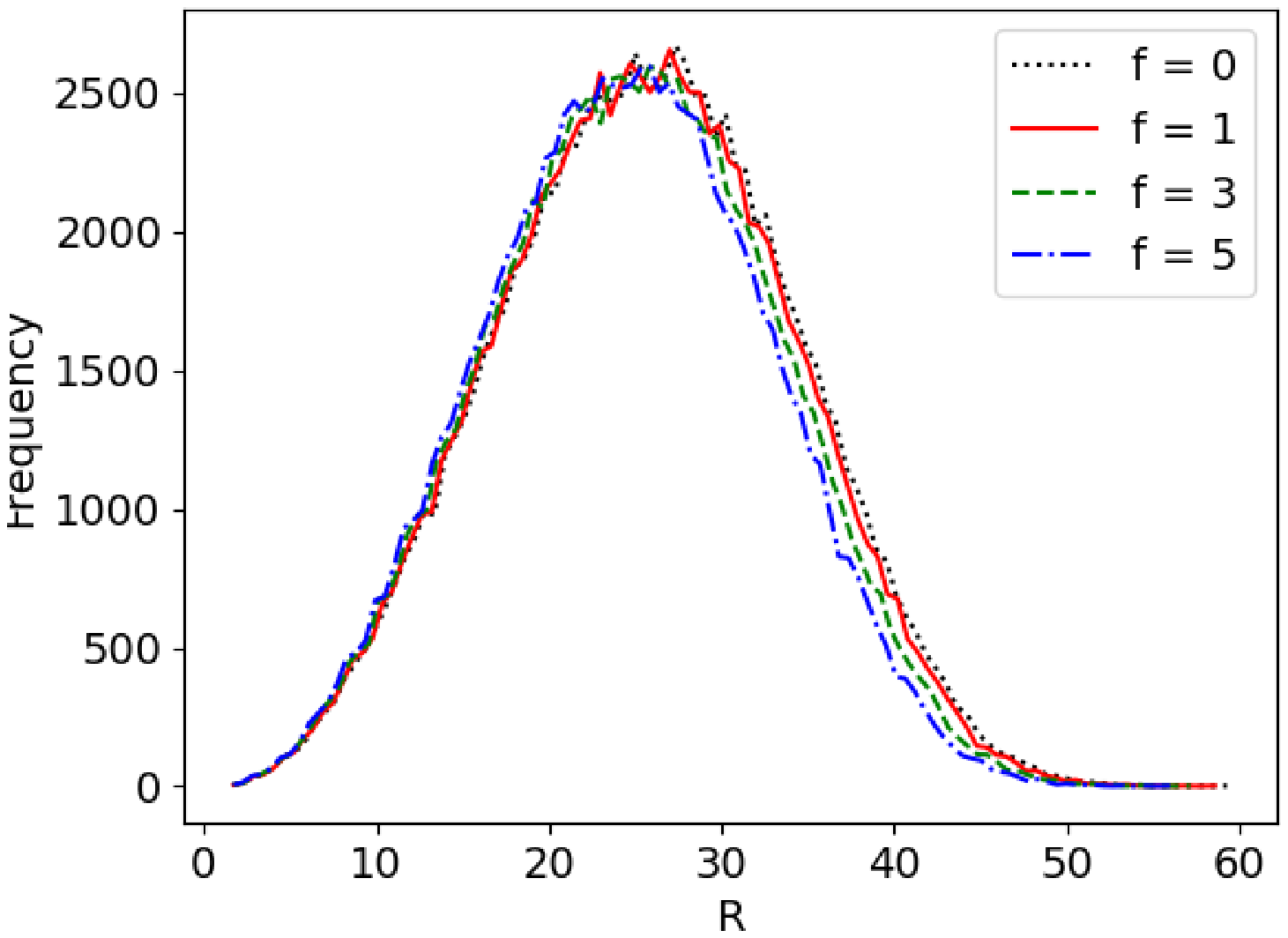}
  \end{center}
  %\vspace{35mm}
  \caption{Simulation results of probability density function of $R$ when $m=100$ and $\alpha=2$}
  \label{fig:alpha2-domain100-stopfailure}
\end{figure}

\begin{figure}[p]
  \vspace{-35mm}
  \begin{center}
    \includegraphics[width=80mm]{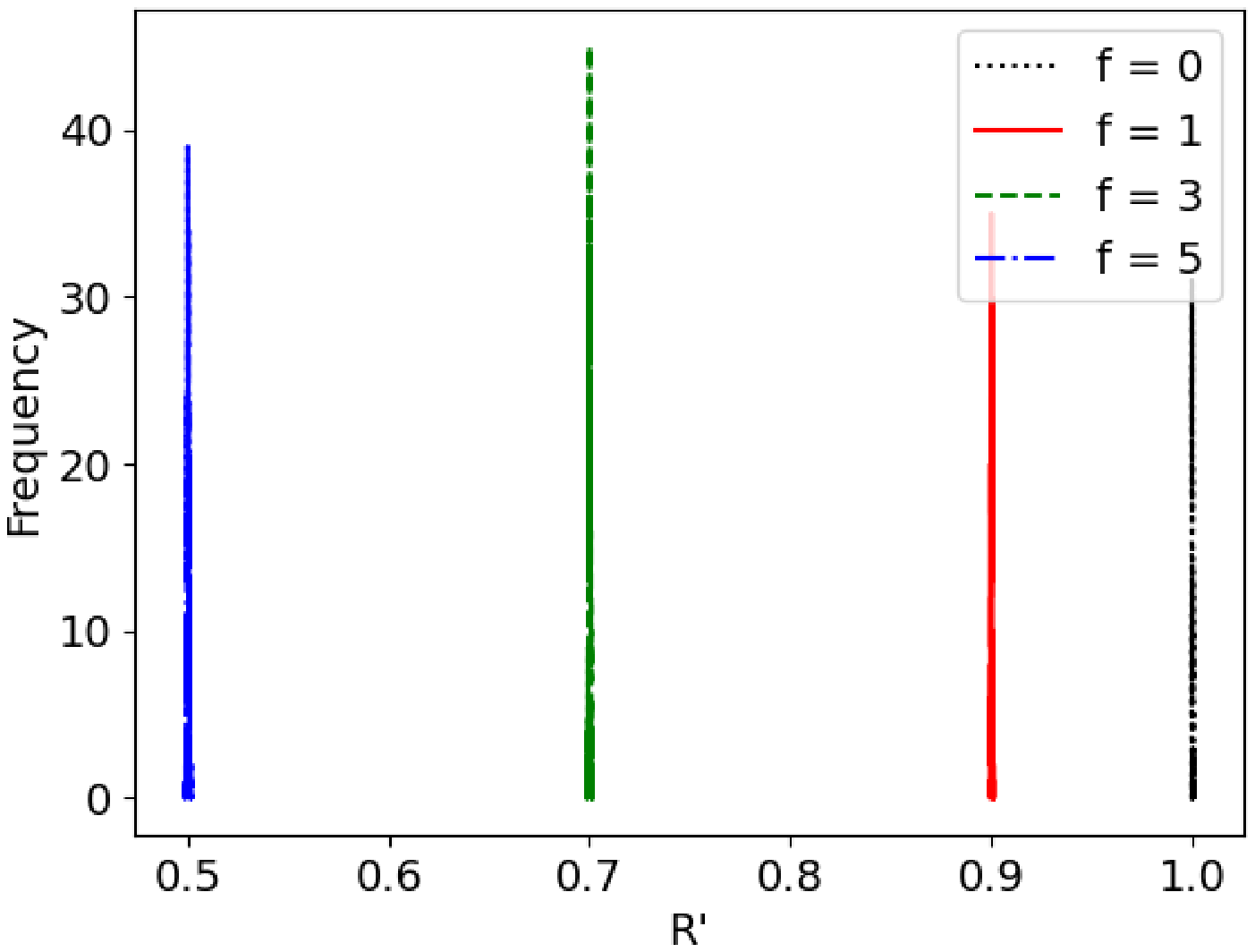}
  \end{center}
  %\vspace{35mm}
  \caption{Simulation results of probability density function of $R'$ when $m=10$, $X=10$\%, and $\alpha=2$}
  \label{fig:alpha2-domain10-0.1-stopfailure}

\vspace{10mm}
  \begin{center}
    \includegraphics[width=80mm]{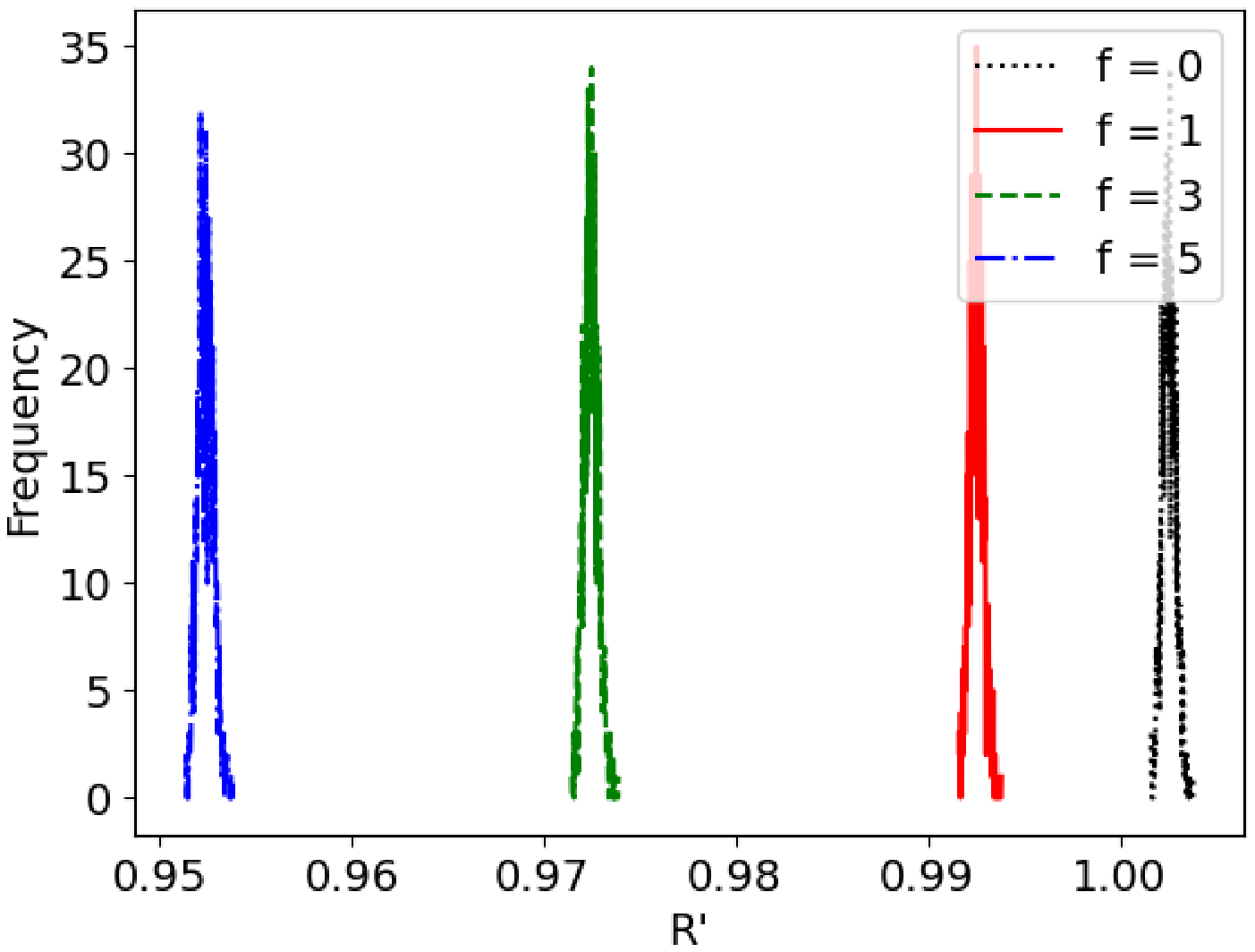}
  \end{center}
  %\vspace{35mm}
  \caption{Simulation results of probability density function of $R'$ when $m=100$, $X=10$\%, and $\alpha=2$}
  \label{fig:alpha2-domain100-0.1-stopfailure}
\end{figure}

\begin{figure}[p]
  \vspace{-35mm}
  \begin{center}
    \includegraphics[width=80mm]{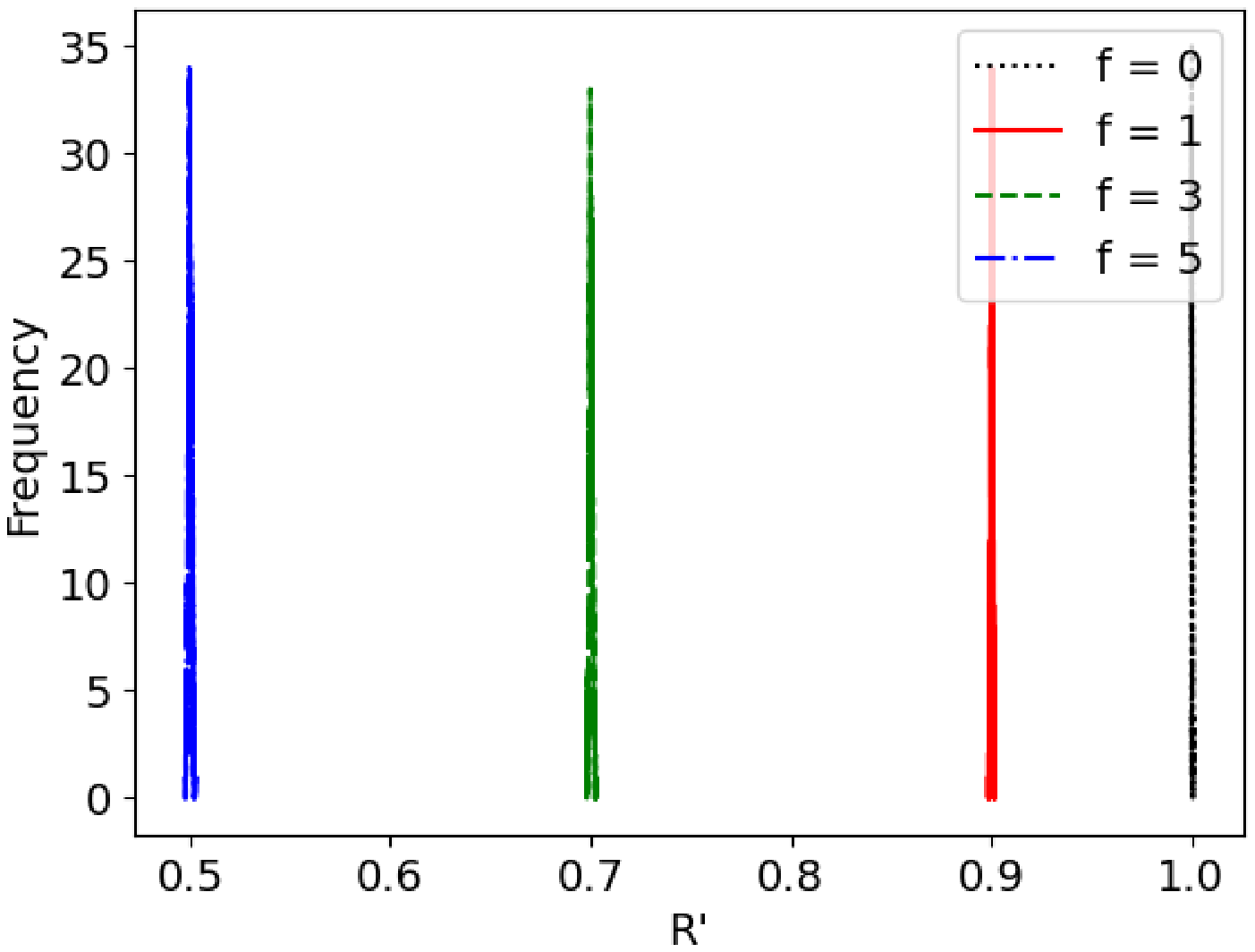}
  \end{center}
  %\vspace{35mm}
  \caption{Simulation results of probability density function of $R'$ when $m=10$, $X=30$\%, and $\alpha=2$}
  \label{fig:alpha2-domain10-0.3-stopfailure}
  \vspace{10mm}
  \begin{center}
    \includegraphics[width=80mm]{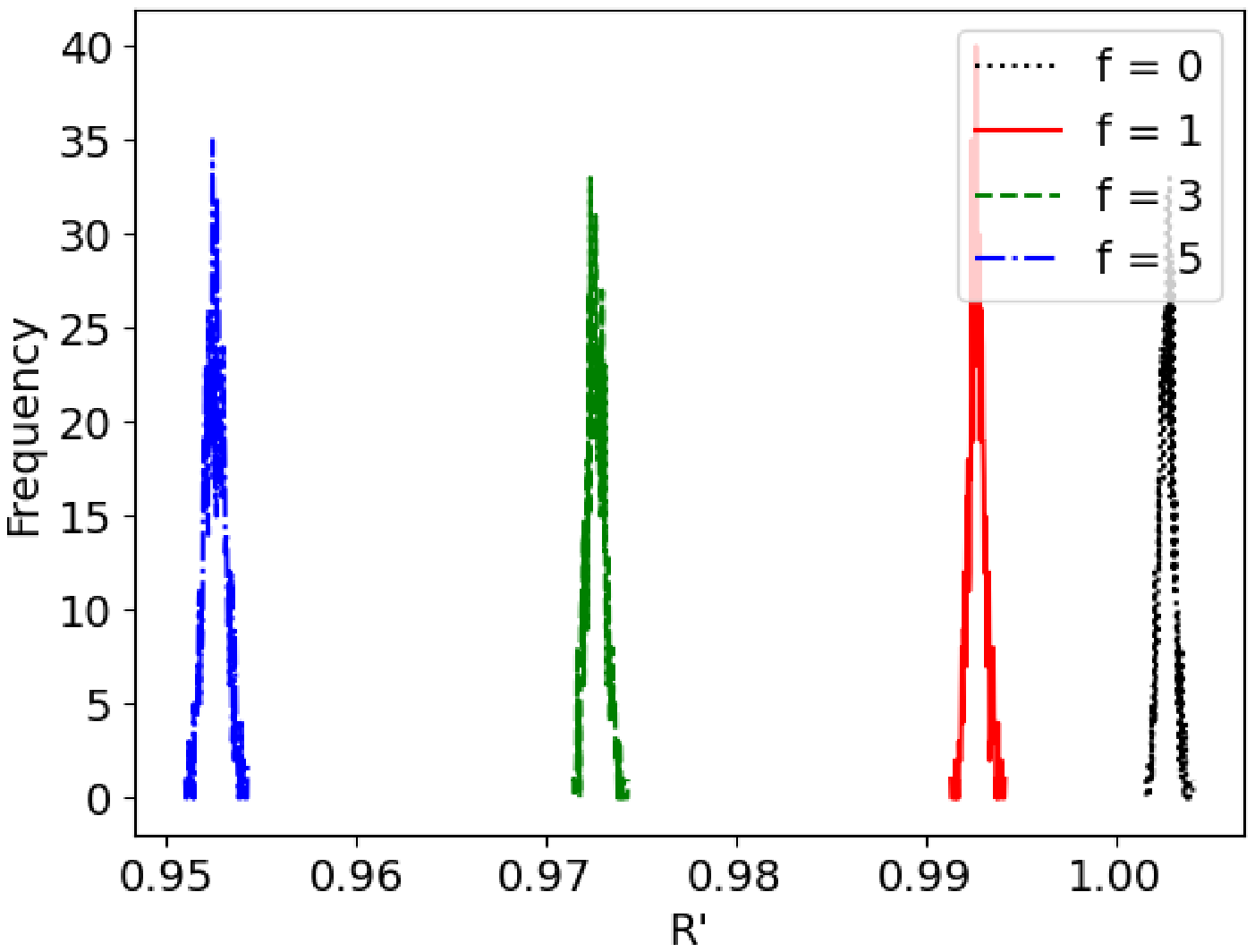}
  \end{center}
  %\vspace{35mm}
  \caption{Simulation results of probability density function of $R'$ when $m=100$, $X=30$\%, and $\alpha=2$}
  \label{fig:alpha2-domain100-0.3-stopfailure}
\end{figure}

 We also estimated the effect of stop failures on tamper resistance in the 
entire system. 
 We show Monte Carlo simulation results when CCNs in $f=1,3,5$ domains 
become stop failure. 
As shown in Figs.~\ref{fig:alpha2-domain10-stopfailure}, \ref{fig:alpha2-domain100-stopfailure},
\ref{fig:alpha2-domain10-0.1-stopfailure}, \ref{fig:alpha2-domain100-0.1-stopfailure}, \ref{fig:alpha2-domain10-0.3-stopfailure} and
\ref{fig:alpha2-domain100-0.3-stopfailure}, the number of failed CCNs is sufficiently large (f = 3 or 5), the tamper resistance 
in the entire system deteriorates significantly.
However, $R$ and $R'$ shifts higher as m increases indicating that the effect of stop failures on tamper resistance is relatively small.
If failure occurs, the $R$ improved as well. 
As $f$ increases, the peak of the histograms in $R$ generally shifts 
toward a higher position. 

However, $R'$ did not improve 
compared with $R$. 
 If $f$ is small and the number of stop-failed CCNs is sufficiently large, 
the tamper resistance in the entire system can degrade significantly. 
 However, as $m$ increases, $R'$ shifts toward the 
higher position, and the effects on tamper resistance can be relatively small. 

Each domain has a demand to strengthen tamper-resistance by joining the 
system to execute our cross-referencing method. 
 Therefore, CCNs with stop failure are assumed to be repaired quickly. 
It is reasonable to assume that only a small number of CCNs experience 
stop failure. 

%

%上に移動
% Each domain has a demand to strengthen tamper-resistance by joining the 
%system to execute our cross-referencing method. 
% Therefore, CCNs with stop failure are assumed to be repaired quickly. 
%It is reasonable to assume that only a small number of CCNs experience 
%stop failure. 

\newpage
\section{Conclusion}
\label{sec:conclusion}

 We explained our previously proposed cross-referencing method for enabling multiple domains of P2P 
networks to manage their own blockchains and periodically exchange among each 
other the state of the latest confirmed block in the blockchain with the 
hysteresis signatures. 
 We also discussed the design of a communication protocol that autonomously executes our 
cross-referencing method among the domains. 

 We theoretically evaluated the effectiveness of our method 
from three viewpoints: decentralization, scalability, and 
tamper-resistance. 
 The method improved the scalability and tamper-resistance 
of a blockchain while reducing the degradation of decentralization. 
 With our method, in-domain decentralization is equivalent to a usual public 
blockchain, such as Bitcoin, while out-domain decentralization is preserved by 
hysteresis signatures, which affects some blocks. 

 We theoretically formulated the Bitcoin scalability problem by estimating 
the transaction processing capacity per second $G(\tau)$. 
 We found that $G(\tau_{fork})$ is the optimal upper bound of 
transaction processing capacity in Bitcoin-type blockchain systems, which 
is the cause of the scalability problem. 
 We demonstrated that our cross-referencing method can break through the limit of 
the upper bound and capacity can reach the same level as that of a the credit card company VISA, Ltd. in theory. 
It is highly possible that the transaction processing capacity will exceed
that of VISA credit card system, but the actual performance
needs to be evaluated experimentally, which is left for our future work. 
 The effectiveness of our method was examined from the theoretical 
perspective by defining two tamper-resistance-improvement ratios $R$ and $R'$. 
 As the number of domains increases, the peaks of the distributions 
of $R$ and $R'$ generally shift toward the higher position.
 We confirmed that the dispersion of this distribution increases as the number 
of domains increases.
 We assumed that the hash rate obeys a Pareto distribution, and
the comparison of the scale parameter of the Pareto distribution $\alpha=2,3$
showed that as $\alpha$ decreases, the peak of the distribution of $R$ shifts 
to the smaller position.
The proposed system of 1,000 domains are 3-10 times more 
tamper-resistant than that of 100 domains, and the capacity is 10 times higher. 
 We also estimated the effect of stop failures on tamper resistance in the 
entire system. 
 We showed that $R$ improves but $R'$ does not. 
However, as the number of domains increases, the peak of the histograms in 
$R$ and $R'$ shift toward the higher position, 
so performance degradation 
due to stop failures is relatively small. 

 We are currently developing a program of CCNs for conducting experimental 
evaluations on our cross-referencing method as a reference implementation of 
the communication protocol between CCNs. 
 The program is open to the public at our Github website \cite{crossrefbc}. 
We will present experimental results elsewhere. 

 For future work, we will determine if our cross-referencing method 
can be successfully applied to various situations, for example, when the 
starting CCN for cross-referencing is not fixed and frequently changes and 
when multiple cross-referencing requests are sent from multiple domains 
at the same time. 
 It is also important to consider security issues in Layer 0 where CCNs can 
share block information securely with each other. 

 This work was partially supported by the Japan Society for the Promotion of 
Science (JSPS) through KAKENHI (Grants-in-Aid for Scientific Research) Grant 
Numbers 17H01742 and 20K11797.

\end{document}